\documentclass{elsart3}

\usepackage{graphicx}

\usepackage{amsmath}
\usepackage{cite}

\begin{document}

\begin{frontmatter}




\title{Compton Scattering by the Nucleon\thanksref{t1}}
\thanks[t1]{Supported by Deutsche Forschungsgemeinschaft SPP(1034) and
  projects SCHU222 and 436 RUS 113/510.
Presented on P.A. Cherenkov and Modern Physics, Moscow, June 22 - 25, 2004}

\author[a1]{Martin Schumacher}
\ead{mschuma3@gwdg.de}

\address[a1]{Zweites Physikalisches Institut der Universit\"at G\"ottingen,\\
   Friedrich-Hund-Platz 1, D-37077 G\"ottingen, Germany}

\begin{abstract}
The status of Compton scattering by the nucleon at energies of the first and
second resonance is summarized. In addition to a general test of dispersion
theories and a precise determination of polarizabilities, the validities of
four fundamental sum rules are explored. {\it Recommended} averages of
experimental values for the electromagnetic polarizabilities and
spin-polarizabilities of the nucleon are determined: $\alpha_p=12.0\pm 0.6$,
$\beta_p=1.9\mp 0.6$,  $\alpha_n=12.5\pm 1.7$, $\beta_n=2.7\mp 1.8$     
(unit $10^{-4}$fm$^3$), $\gamma^{(p)}_\pi=-38.7\pm 1.8$,
$\gamma^{(n)}_\pi=58.6\pm 4.0$ (unit $10^{-4}$ fm$^4$).
\end{abstract}


\end{frontmatter}

\section{Introduction}
\label{sec1}
One of the central challenges of hadron physics in the regime of strong
(non-perturbative) QCD is to identify the relevant degrees of freedom of the
nucleon and to quantitatively explain experimental data in terms of these
degrees of freedom. Among the processes studied so far Compton scattering
plays a prominent role because of the well understood properties of the
electromagnetic interaction. Compton scattering is parameterized in terms of
generalized polarizabilities of which the invariant amplitudes given in
\cite{lvov97,wissmann03} are the most convenient ones. Scattering proceeds 
through Born terms
and through excitation of internal degrees of freedom of the nucleon, the
latter leading to the electromagnetic polarizabilities $\alpha$ and $\beta$ 
and to the spin polarizabilities $\gamma_0$ and $\gamma_\pi$. These 
quantities may be understood as low-energy limits of
invariant amplitudes \cite{lvov97,wissmann03,schumacher03}.  
Exchanges in the $t$-channel are
responsible for the validities of four sum rules, {\it viz.} the Baldin or
Baldin-Lapidus \cite{baldin60} (BL) sum rule for $(\alpha+\beta)$, the
Gerasimov-Drell-Hearn \cite{gerasimov66} (GDH) sum rule for $\kappa^2$
($\kappa$ being the anomalous magnetic moment), the
Bernabeu-Ericson-FerroFontan-Tarrach \cite{bernabeu74} (BEFT) sum rule for
$(\alpha-\beta)$ and the Lvov-Nathan \cite{lvov99} (LN) sum rule for 
$\gamma_\pi$.
A test of these sum rules, therefore, contains valuable information on the
physical properties of these exchanges. The $t$-channel exchanges $\pi^0$ and 
$\sigma$ dominate the sizes of $\gamma_\pi$ and $(\alpha-\beta)$,
respectively, and, therefore, have to be included into the model of the
nucleon
\cite{levchuk05}.

\section{New Results}
\label{sec2}

\subsection{Electromagnetic polarizabilities and BL sum rule}

Precise experimental values for the polarizabilities of the proton 
are available through a
recent reevaluation of all low-energy Compton scattering experiments carried
out in the 1950's to 1990's \cite{baranov01} and through a recent experiment 
carried
out at MAMI (Mainz) \cite{olmos01}. The  results are listed in lines 2 
and 3 of
Table 1. The two results for the sum $\alpha+\beta$ of 
\begin{table}[h]
\caption{Electromagnetic polarizabilities of the proton.}
\begin{center}
\begin{tabular}{|c|c|c|c|}
\hline
$\alpha_p$& $\beta_p$ & $\alpha_p+\beta_p$&\\
\hline
$11.7 \pm 1.1$ & $ 2.3 \pm 1.1$ & $14.0\pm 1.6$& \cite{baranov01}\\
$11.9 \pm 1.4$ & $ 1.2 \pm 0.8$ & $13.1\pm 1.6$& \cite{olmos01}\\ 
\hline
$11.8 \pm 0.9$ & $ 1.6 \pm 0.6$ & $13.6\pm 1.1$& a)\\
$12.0 \pm 0.6$ & $ 1.9 \mp 0.6$ & $13.9\pm 0.3$& b)\\ 
\hline
\end{tabular}
\end{center}
a) weighted average of the data in lines 2 and 3\\
b) the same as a) but constrained by the BL sum rule \cite{olmos01,levchuk00}
\label{table1}
\end{table}
electromagnetic polarizabilities listed in lines 2 and 3 have a weighted
average of $\alpha+\beta=13.6\pm 1.1$ which has to be compared with the BL
sum-rule prediction  
\begin{equation}
\alpha+\beta=\frac{1}{2\pi^2}\int^\infty_{\omega_0}\frac{\sigma_{\rm
    tot}(\omega)}{\omega^2}d\omega
\label{equation1}
\end{equation}
obtained from photo-absorption data, {\it viz.} $13.9\pm 0.3$ 
\cite{olmos01,levchuk00}. The excellent agreement of the two results
may be considered as a precise confirmation of the BL sum rule. 

For the electromagnetic polarizabilities of the neutron three different methods
have led to results of good precision. These are the electromagnetic
scattering of slow neutrons in the Coulomb field of heavy nuclei 
\cite{schmiedmayer91},
the quasi-free Compton scattering on neutrons initially bound in the deuteron
\cite{kossert02} and the coherent-elastic scattering of photons by the deuteron
\cite{lundin03}. The results are given in Table 2. Since the model 
dependence of the
method of coherent-elastic scattering is still under discussion we recommend
to base the average only on the two other results.
\begin{table}[h]
\caption{Electromagnetic polarizabilities of the neutron determined by three
  different methods and their {\it recommended} average.}
\begin{tabular}{|l|l|}
\hline
experimental method& polarizability\\
\hline
electromagnetic scattering \cite{schmiedmayer91}& $\alpha_n= 12.6 \pm 2.5$\\   
quasi-free Compton scattering \cite{kossert02}& $\alpha_n= 12.5 \pm 2.3$\\
coherent-elastic scattering \cite{lundin03}& $\alpha_n= 8.8 \pm 3.8$\\
\hline
{\it recommended} average with $\beta_n$ & $\alpha_n= 12.5 \pm 1.7$\\
constrained by $\alpha_n+\beta_n= 15.2\pm 0.5$ \cite{levchuk00}& 
$\beta_n=2.7\mp 1.8$\\
\hline
\end{tabular}
\label{table2}
\end{table}

\subsection{The GDH sum rule}

The integrand of the GDH sum rule
\begin{equation}
\frac{2\pi^2 \alpha_e \kappa^2}{M^2}= \int^\infty_{\omega_0} 
\frac{\sigma_{3/2}(\omega) -\sigma_{1/2}(\omega)}{\omega}d\omega
\label{equation2}
\end{equation}
with $\alpha_e=1/137.04$ and $\sigma_{3/2}(\omega)$ and  
$\sigma_{1/2}(\omega)$ the photo-absorption cross sections for parallel and 
antiparallel nucleon and photon spins, respectively, was determined at two
electron accelerators using hydrogen targets. While the measurement 
from 0.2 to 0.8 GeV was
carried out at MAMI \cite{ahrens00}, the energy range from 0.68 to 2.9 GeV was
covered at the electron stretcher ring ELSA \cite{dutz03}. 
\begin{table}[h]
\caption{Measured values of the GDH integral $I_{\rm GDH}$ for the proton
and model predictions for the unmeasured ranges.}
\begin{tabular}{|l|c|c|}
\hline
 & $E_\gamma$ [GeV] & $I_{\rm GDH} [\mu b]$\\
\hline
MAID2002 \cite{tiator02} &$0.14 - 0.20$& $-27.5 \pm 3$\\
measured (GDH-Collaboration)& $0.20 - 2.90$ & $254\pm 5 \pm 12$\\
Regge prediction \cite{bianchi99}&$>$ 2.9& $-14$\\
\hline
GDH integral & $0.14 - \infty$& $\approx 213 $\\
\hline
GDH sum rule &  $\nu_{thr} - \infty$& $205 $\\
\hline
\end{tabular}
\label{table3}
\end{table}
In   total, the  range from
the resonance region up to the Regge regime  was covered. This range is wide
enough to reliably make conclusions on the validity of the GDH sum rule for
the first time. Only the ranges from 0.14 - 0.20 GeV and $>$ 2.9 GeV had to be
covered using model predictions.

\subsection{Spin polarizability $\gamma_\pi$ and LN sum rule}

The LN sum rule \cite{lvov99} is formulated for the backward
direction using fixed-$\theta$ dispersion theory. 
Its $s$-channel part is given by
\begin{eqnarray}
&& \gamma^s_\pi=\int^\infty_{\omega_0} \frac{d\omega}{4\pi^2\omega^3}
\sqrt{1+\frac{2\omega}{M}} \left(1+\frac{\omega}{M}\right)\nonumber\\
&&\quad\quad\quad\times\sum_n 
P_n[\sigma^n_{3/2}(\omega)-\sigma^n_{1/2}(\omega)]
\end{eqnarray}
with  $P_n = +1$ for $M1,E2,M3,\cdots$ multipoles and $P_n = -1$ for
$E1,M2,E3,\cdots$ multipoles. Its $t$-channel part is given by the 
contributions of pseudoscalar poles
\begin{eqnarray}
&&\gamma^t_\pi=\frac{1}{2\pi M}
\Big[ \frac{g_{\pi NN}F_{\pi^0\gamma\gamma}}{m^2_{\pi^0}}\tau_3
+ \frac{g_{\eta NN}F_{\eta\gamma\gamma}}{m^2_{\eta}}\nonumber\\
&&\hspace{4cm}
+ \frac{g_{\eta' NN}F_{\eta'\gamma\gamma}}{m^2_{\eta'}} \Big].
\label{equation4}
\end{eqnarray}
The analogous separation in fixed-$t$ dispersion
theory into an integral part and an asymptotic (contour integral) part
is described in \cite{lvov97}. In the present approach  we 
tentatively make the 
assumption that  $\gamma^t\equiv \gamma^{\rm as}$.

The second line in Table \ref{table4} shows
\begin{table}[h]
\caption{Experimental results obtained for the backward spin-polarizabilities
  of proton and neutron compared with predictions from the LN sum rule. Unit 
of spin polarizability: $10^{-4}{\rm fm}^4$.}
\begin{tabular}{|l|l|l|l|}
\hline
spin polarizabil.   & proton     &neutron   &\\
\hline
$\gamma_\pi$(fixed-$t$) & $-38.7\pm 1.8$& $+58.6\pm 4.0$ & exp. 
\cite{kossert02,camen02}\\
$\gamma_\pi$(fixed-$\theta$)& $-39.5\pm 2.4$& $+52.5\pm 2.4$& sum rule
\cite{lvov99}\\
\hline
$\gamma^t\equiv \gamma^{\rm as}$& $-46.6$& $+43.4$&$\pi^0 + \eta + \eta'$
\cite{lvov99} \\
$\gamma^{\rm int}_\pi$ & $+7.9 \pm 1.8$ & $+15.2\pm 4.0$& exp. 
\cite{kossert02,camen02}\\
$\gamma^s_\pi$&$ +7.1\pm 1.8$ & $+9.1 \pm 1.8$ & sum rule \cite{lvov99}\\
\hline
\end{tabular}
\label{table4}
\end{table}
the spin polarizabilities determined from experimental data 
\cite{kossert02,camen02} using
fixed-$t$ dispersion theory for the data analysis.  Apparently,
for the proton the integral part $\gamma^{\rm int}$ of fixed-$t$ dispersion
theory (line 5) and the $s$-channel part $\gamma^s$ of fixed-$\theta=\pi$ 
dispersion theory (line 6)
are in agreement with each other with good precision. The same is true
for the neutron within a larger margin of uncertainty. This leads to the
important conclusion that the assumption $\gamma^t_\pi\equiv \gamma^{\rm as}$
is confirmed and that the backward spin-polarizability is fully understood.

\subsection{The electromagnetic polarizability $(\alpha - \beta)$ and  
the BEFT sum rule}

The $s$-channel part of the BEFT sum rule \cite{bernabeu74} is given by
\begin{eqnarray}
&&(\alpha-\beta)^s=\frac{1}{2\pi^2}\int^\infty_{\omega_0}
\frac{d\omega}{\omega^2}\sqrt{1+\frac{2\omega}{M}}\nonumber\\
&&\quad\quad\quad \times 
\left[\sigma(\Delta P={\rm yes})-\sigma(\Delta P={\rm no})
\right]
\label{equation5}
\end{eqnarray}
where $\Delta P$ = yes for $E1,M2,E3,\cdots$ multipoles and $\Delta P$ = no
for $M1,E2,E3,\cdots$ multipoles. The $t$-channel part of the BEFT sum rule
is given by
\begin{align}
&(\alpha-\beta)^t=\frac{1}{16\pi^2}\int^\infty_{4m^2_\pi}\frac{dt}{t^2}
\frac{16}{4M^2-t}\left( \frac{t-4m^2_\pi}{t}\right)^{1/2}\nonumber\\
&\left[f^0_+(t)F^{0*}_0(t)-\left(M^2-\frac{t}{4}\right)\left(\frac{t}{4}
-m^2_\pi\right)f^2_+(t)F^{2*}_0(t)\right],
\label{equation6}
\end{align}
where $f^J_+(t)$ ($J$ = 0, 2)  are the amplitudes for the 
$\pi\pi\to N\bar{N}$
transition and $F^J_0(t)$ the amplitudes for the $\gamma\gamma\to \pi\pi$
transition. Both amplitudes are constructed from the respective imaginary
parts through dispersion relations ($N/D$ method) which incorporate the effects
of $\pi\pi$ correlations as given by the phase function $\delta^J_0(t)$
extracted from $\pi\pi$ scattering $(\pi N\to N\pi\pi)$ data. 
It is of interest to relate the procedure outlined above to the $\sigma$ meson.
 The $\sigma$ meson -- though strongly demanded by
theory as a chiral partner of the $\pi$ meson -- has only recently been
observed in other experimental data. It has been realized 
\cite{colangelo01,ishida03}
that the phase shift $\delta^0_0(t)$ in scalar (J=0) and iso-scalar (I=0)
$\pi\pi$ scattering data can consistently be interpreted in terms of a broad
resonance if a background is taken into account. This background shifts the
position of $\sqrt{s}(\delta_S=90^\circ)$ from  Re$\sqrt{s}$(pole) to about
900 MeV. This leads to the results listed in lines 2 and 3 of Table
\ref{table5}.
\begin{table}[h]
\caption{Position of the $\sigma(600)$ pole, $90^\circ$ crossing of the scalar
  phase and two-photon decay width of the $\sigma$ meson, with supplement a)
given by the present author.}
\begin{tabular}{|c|c|c|}
\hline
$\sqrt{s}$(pole) [MeV] &   $\sqrt{s}(\delta_S=90^\circ)$& method\\
\hline
$(470\pm 30) -i (295\pm 20)$&$844 \pm 13$ MeV & $\pi N\to N\pi\pi$ 
\cite{colangelo01}\\
$(585\pm 20) -i (193\pm 35)$&$\sim 900$ MeV & $\pi N\to N\pi\pi$ 
\cite{ishida03}\\
\hline
\hline
$\sqrt{s}$(pole) [MeV] & $\Gamma_{\sigma\to 2\gamma}$ [keV]& \\
\hline
$(547\pm 45)-i(602\pm 181)$& $0.62\pm 0.19$ & $\gamma\gamma\to\pi^0\pi^0$
\cite{filkov99}\\
\hline
PDG summary$^{a)}$&& {\it recommended}\\
$(500 \pm 40) - i ( 250\pm 40)$&& \!average \cite{eidelman04}\\
\hline
\end{tabular}
{\footnotesize a) supplemented by estimated errors.}
\label{table5}
\end{table} 
Line 5 of Table 5 shows the results of an analysis of the $\gamma\gamma\to
\pi^0\pi^0$ reaction and line 7 the PDG summary on the pole parameters.
\begin{table}[h]
\caption{Numerical evaluation of the BEFT sum rule, with corrections a) and b)
supplemented by the present author.}
\begin{center}
\begin{tabular}{|l|l|l|c|}
\hline
$(\alpha_p-\beta_p)^s$& $(\alpha_p-\beta_p)^t$&$(\alpha_p-\beta_p)^{\rm
  BEFT}$& authors\\
\hline
\,$-$4.92 &\,+9.28$^{a)}$&\,+4.36&\cite{guiasu78}\\
\,$-$4 &\,+10.4$^{b)}$&\,+6.4 &\cite{budnev79}\\
\,$-$5.42&\,+8.6&\,$+(3.2^{+2.4}_{-3.6})$&\cite{holstein94}\\
\,$-$5.56&\,+16.46&\, $+(10.7 \pm 0.2)^{c)}$& \cite{drechsel03}\\
\hline
\,$-(5.0\pm 1.0)$&\,$+(14.0\pm 2.0)$&\,$+(9.0\pm 2.2)$&\cite{levchuk04}\\
\hline
\end{tabular}
\end{center}
{\footnotesize a) correction for the D wave contribution (-1.7) included. b)
correction for the polarizability of the pion (+3.0) included. c) best value
from a range of results given by the authors.}
\label{table6}
\end{table}
Table \ref{table6} shows five independent evaluations of the BEFT sum rule,
where the last one corresponds to recent 
unpublished work \cite{levchuk04}. 
The $s$-channel
component of this last evaluation $(\alpha_p-\beta_p)^s=-(5.0\pm 1.0)$ is
in line with all the other results, whereas the $t$-channel result
$(\alpha_p-\beta_p)= +(14.0\pm 2.0)$ is in agreement with the result of
\cite{drechsel03} only.
Adopting  the results of line 6 
we arrive at the conclusion
that the prediction of the BEFT sum rule, {\it viz.} 
$(\alpha_p-\beta_p)^{\rm BEFT}= +(9.0\pm 2.2)$, is in line 
with the experimental value 
$(\alpha_p-\beta_p)^{\rm exp}=10.1\pm 0.9$ (see Table \ref{table7}).  
This result has to be compared with the expectation that the $\sigma$
meson probably has a two-component structure with  $|\pi\pi\rangle$
being the one component and $|q\bar{q}\rangle$ being the other
\cite{scadron04}.

As a summary Table \ref{table7}
\begin{table}[h]
\caption{Predicted information $(\alpha-\beta)^{\rm calc}_p$ on the
  polarizability difference compared with the experimental result
  $(\alpha-\beta)^{\rm exp}_p=10.5\pm 1.1$ \cite{olmos01} or
$(\alpha-\beta)^{\rm exp}_p=10.1\pm 0.9$ (adopted average including all
  existing data \cite{baranov01,olmos01}). In fixed-$t$ dispersion theory the
  prediction $(\alpha-\beta)^{\rm calc}_p$ corresponds to the integral part 
$(\alpha-\beta)^{\rm int}_p$, in fixed-$\theta$ dispersion theory 
 $(\alpha-\beta)^{\rm calc}_p$ is either chosen to be the $s$-channel
  contribution only (line 3) or the predicted $s$-channel contribution 
supplemented 
by the predicted $\gamma\gamma\to\pi\pi\to N\bar{N}$ $t$-channel contribution
  according to the BEFT sum rule (line 4).}
\begin{tabular}{|l|l|l|}
\hline
theor.&  $(\alpha-\beta)^{\rm calc}_p$&  
$(\alpha-\beta)^{\rm exp}_p$ - $(\alpha-\beta)^{\rm calc}_p$\\
\hline
fixed-$t$ & $- 3.1$ (int. part) \cite{wissmann03}
& $(\alpha-\beta)^{\rm as}_p=13.2\pm 1.3$\\
fixed-$\theta$ & $-(5.0\pm 1.0)$ ($s$-chan.)& 
$(\alpha-\beta)^{t\mbox{-}{\rm exp}}_p=15.1\pm 1.3$\\    
fixed-$\theta$ & $+(9.0\pm 2.2)$ ($s+t$-ch.)& 
$(\alpha-\beta)^{t \mbox{-}{\rm miss}}=1.1  \pm 2.4$\\
\hline
\end{tabular}
\label{table7}
\end{table}
shows the difference between calculated values for $(\alpha-\beta)_p$ and the
experimental value for this quantity obtained under conditions explained in
the caption of the table. We see that in fixed-$t$ dispersion theory we have
to explain $(\alpha-\beta)^{\rm as}_p= 13.2\pm 1.3$ through a contribution
which has no interpretation in terms of the  integral part in fixed-$t$
dispersion theory. This value, therefore, may be interpreted as an empirical
result for the asymptotic contribution to $(\alpha-\beta)$. 
In case of 
fixed-$\theta$ dispersion theory 
$(\alpha-\beta)^{t \mbox{-}{\rm miss}}_p$ is compatible with zero.

An interesting alternative for the prediction of $(\alpha-\beta)$ 
which deserves further consideration has been proposed and evaluated by
Akhmedov and Fil'kov \cite{akhmedov81}. In this approach $(\alpha-\beta)$
is expressed through a dispersion relation at fixed $u=M^2$ in the point 
$t=0$.

\subsection{The effective $\sigma$ pole}

The evaluation of the reaction $\gamma\gamma\to\pi^0\pi^0$ \cite{filkov99}
has led to a determination of the position of the $\sigma$ pole as well as to
the determination of the photon decay width $\Gamma_{\sigma\to 2\gamma}$ 
(Table \ref{table5}). This makes it interesting to reinvestigate the
properties of the effective $\sigma$ pole introduced to approximately
represent the asymptotic (contour-integral) contribution \cite{lvov97},
using these data as a basis. The relation between $\alpha-\beta$ and the
$\sigma$ pole is given by
\begin{equation}
\frac{g_{\sigma NN}F_{\sigma\gamma\gamma}}{2\pi m^2_{\sigma}}=
(\alpha-\beta)^{\sigma \mbox{-} {\rm pole}}
\label{equation7}
\end{equation}
where $m_{\sigma}$ = 600 MeV $\hat{=}$ 3.04 fm$^{-1}$ is the nominal value of
the $\sigma$ mass which has been found to be consistent with experimental
Compton differential cross-sections in the second resonance 
region \cite{galler01}. The relation to the decay width is given by
\begin{equation}
g_{\sigma NN}F_{\sigma\gamma\gamma}=+16 \pi \sqrt{\frac{g^2_{\sigma NN}}{
4\pi}\frac{\Gamma_{\sigma\to 2\gamma}}{m^3_\sigma}}
\label{equation8}
\end{equation}
where the approximate equality of the two coupling constants
\begin{equation}
\frac{g^2_{\sigma NN}}{4\pi}\simeq  \frac{g^2_{\pi NN}}{4\pi}=13.75
\label{equation9}
\end{equation}
may be justified through the linear $\sigma$ model. Using 
\begin{equation}
\Gamma_{\sigma\to 2\gamma}=(0.62\pm 0.19)\rm {\rm keV}
\label{equation10}
\end{equation}
evaluated by Fil'kov and Kashevarov \cite{filkov99}, we arrive at
\begin{equation}
(\alpha-\beta)^{\sigma \mbox{-}{\rm pole}}= 10.7 \pm 1.7.
\label{equation11}
\end{equation}
The same calculation may be carried out using the experimental value
$m_\sigma=(547 \pm 45)$ MeV as obtained by Fil'kov and Kashevarov
\cite{filkov99}. Then we arrive at
\begin{equation}
(\alpha-\beta)^{\sigma \mbox{-}{\rm pole}}= 14.8^{+2.1}_{-2.5}
(\Delta\Gamma_{\sigma\to 2\gamma}){}^{+5.2}_{-3.6}(\Delta m_\sigma).
\label{equation12}
\end{equation}
It is satisfactory to see that the
numbers obtained in (\ref{equation11}) and (\ref{equation12})
for $(\alpha-\beta)^{\sigma \mbox{-}{\rm pole}}$ are in the same 
range as the number given for $(\alpha -\beta)^{\rm as}_p$
in the third row of Table \ref{table7}.
 \\
{\footnotesize Acknowledgement: The author is indebted to M.I. Levchuk,
  A.I. L'vov and A.I. Milstein for valuable discussions.}


\begin{thebibliography}{}

\bibitem{lvov97}
A.I. L'vov, V.A. Petrun'kin, M. Schumacher, Phys. Rev. C 55 (1997) 359--377;

\bibitem{wissmann03}
F. Wissmann, SPRINGER TRACTS IN MODERN PHYSICS 200 (2003) 1--132

\bibitem{schumacher03}
M. Schumacher, Nucl. Phys. A 721 (2003) 773c--776c; nucl-ex/0309010

\bibitem{baldin60}
A.M. Baldin, Nucl. Phys. 18 (1960) 310--317; L.I. Lapidus,
Zh. Eksp. Teor. Fiz. 43 (1962) 1358-1361 [Sov. Phys. JETP 16 (1963) 964--966]

\bibitem{gerasimov66}
S.B. Gerasimov, Sov. J. Nucl. Phys. 2 (1966) 430--433
[J. Nucl. Phys. (U.S.S.R.) 2 (1965) 598--602]; S.D. Drell, A.C. Hearn,
Phys. Rev. Lett. 16 (1966) 908--911

\bibitem{bernabeu74}
J. Bernabeu, T.E.O. Ericson, C. Ferro Fontan, Phys. Lett. 49 B (1974) 381--384;
J. Bernabeu, B. Tarrach, Phys. Lett. 69 B (1977) 484--488

\bibitem{lvov99}
A.I. L'vov, A.M. Nathan, Phys. Rev. C 59 (1999) 1064--1069

\bibitem{levchuk05}
M.I. Levchuk, A.I. L'vov, A.I. Milstein, M. Schumacher, (in preparation)


\bibitem{baranov01}
P.S. Baranov, A.I. L'vov, V.A. Petrun'kin, L.N. Shtarkov, Phys. Part. Nuc. 32
(2001) 376--394

\bibitem{olmos01}
V. Olmos de Le\'on et al., Eur. Phys. J. A 10 (2001) 207--215

\bibitem{levchuk00}
M.I. Levchuk, A.I. L'vov, Nucl. Phys. A 674 (2000) 449--492

\bibitem{schmiedmayer91}
J. Schmiedmayer et al., Phys. Rev. Lett. 66 (1991) 1015--1018

\bibitem{kossert02}
K. Kossert et al., Phys. Rev. Lett. 88 (2002) 162301-1--4; Eur. Phys. J. A 16
(2003) 259--273; nucl.-ex/0210020

\bibitem{lundin03}
M. Lundin et al., Phys. Rev. Lett. 90 (2003) 192501-1--4

\bibitem{ahrens00}
J. Ahrens et al., Phys. Rev. Lett. 84 (2000) 5950--5954; Phys. Rev Lett. 87
(2001) 022003-1--5; Phys. Rev. Lett. 88 (2002) 232002-1--5

\bibitem{dutz03}
H. Dutz et al., Phys. Rev. Lett. 91 (2003) 192001-1--5; Phys. Rev. Lett. 93
(2004) 032003-1--5

\bibitem{tiator02}
L. Tiator, Proc. GDH2002, Genova, Italy 3--6 July, World Scientific (2002)
27--36

\bibitem{bianchi99}
N. Bianchi, E. Thomas, Phys. Lett. B 450 (1999) 439--447;
S. Simula et al., Phys. Rev. D 65 (2002) 034017-1--17

\bibitem{camen02}
M. Camen et al., Phys. Rev. C 65 (2002) 032202-1--3

\bibitem{colangelo01}
G. Colangelo, J. Gasser, H. Leutwyler, Nucl. Phys. B 603 (2001) 125--179

\bibitem{ishida03}
M. Ishida, Prog. Theor. Phys. Suppl. 149 (2003) 190--202; hep-ph/0212383


\bibitem{filkov99}
L.V. Fil'kov, V.L. Kashevarov, Eur. Phys. J. A5 (1999) 285--292

\bibitem{eidelman04}
S. Eidelman et al. (Particle Data Group), Phys. Lett. B 592 (2004) 1;
URL:http://pdg.lbl.gov


\bibitem{guiasu78}
I. Guiasu, E.E. Radescu, Phys. Rev. D 18 (1978) 1728--1730

\bibitem{budnev79}
V.M. Budnev, V.A. Karnakov, Yad. Fiz. 30 (1979) 440--444;[Sov.
J. Nucl. Phys. 30 (1979) 228--230]

\bibitem{holstein94}
B.R. Holstein, A.M. Nathan, Phys. Rev. D 49 (1994) 6101--6108

\bibitem{drechsel03} 
D. Drechsel, B. Pasquini, M. Vanderhaeghen, Phys. Rept. 378 (2003) 99--205

\bibitem{levchuk04}
M.I. Levchuk, A.I. L'vov, A.I. Milstein, M. Schumacher, (to be published)

\bibitem{scadron04}
M.D. Scadron, et al., Phys. Rev. D 69 (2004) 014010;
Erratum-ibid. D 69 (2004) 059901; hep/0309109


\bibitem{akhmedov81}
D.M. Akhmedov, L.V. Fil'kov, Sov. J. Nucl. Phys. 33 (1981) 573--578
[Yad. Fiz. 33 (1981) 1083--1095]

\bibitem{galler01}
G. Galler et al., Phys. Lett. B 503 (2001) 245--255; S. Wolf et al.,
Eur. Phys. J. A 12 (2001) 231-252



\end{thebibliography}
\end{document}